\documentclass[12pt,a4paper]{article}
\def\ba{\begin{array}}
\def\ea{\end{array}}
\def\be{\begin{equation}}
\def\ee{\end{equation}}
\def\lbl{\label}
\def \rf  {(\ref}

\def\x0{\x_0}
\def\x1{\x_1}

\def\real{{\bf R}}
\def\cd{\cal {D}}
\def\cg{{\cal G}}
\def\tcg{\tilde{{\cal G}}}
\begin{document}
\author{L. Hlavat\'y, L. \v Snobl
\thanks{{Email: hlavaty@br.fjfi.cvut.cz}, {snobl@newton.fjfi.cvut.cz}} 
\\ {\it Faculty of Nuclear Sciences and Physical Engineering,} 
\\{ \it Czech Technical University,} 
\\ {\it B\v rehov\'a 7, 115 19 Prague 1, Czech Republic}}

\title{Classification of Poisson--Lie T--dual models with two--dimensional targets}
\date{December 20, 2001}
\bibliographystyle{unsrt}

\maketitle
\abstract{Four--dimensional Manin triples and Drinfeld doubles are classified and corresponding 
two--dimensional Poisson--Lie T--dual sigma models on them  
are constructed. The simplest example of a Drinfeld double 
allowing decomposition into two nontrivially different Manin triples is presented.
}
\section{Introduction}
A very important symmetry of string theories, or more specifically, two--dimensional sigma models is the T--duality. 
In the pioneering work \cite{klse:dna}, Klim\v c\'{\i}k 
and \v Severa introduced its nonabelian version  -- the Poisson--Lie T--duality and showed 
that the dual sigma models can be formulated on Drinfeld doubles. The explicit form of dual models on the nonabelian double 
$GL(2|\real)$ was presented  in the following work \cite{kli:pltd}. Other dual models were given in a series of forthcoming 
papers, see e.g. \cite{sfe:pltd}, \cite{vall:su2}, \cite{mabe:pltqlb}. An attempt to classify all dual principal sigma 
models with three--dimensional target space \cite{jare:pltd} made us to revisit the models with the two--dimensional 
targets and classify them. In the following we classify all four--dimensional Drinfeld doubles and the Poisson--Lie T--dual 
models on them.

\section{Classification of four--dimensional Drinfeld doubles} 
The Drinfeld double $D$ is defined as a Lie group such that its Lie algebra $\cd$ equipped by a symmetric ad--invariant 
nondegenerate bilinear form $\langle .,.\rangle $ can be decomposed into a pair of maximally isotropic 
subalgebras $\cg$, $\tcg$ such that $\cd$ as a vector space is the direct sum of $\cg$ and $\tcg$. Any such decomposition written as 
an ordered set $(\cd,\cg,$$\tcg)$ is called a Manin triple. It is clear that to any Drinfeld double exist at least two Manin triples 
$(\cd,\cg,$$\tcg)$, $(\cd,$$\tcg,\cg)$. Later we show an example of Drinfeld double with more than two possible decomposition into 
Manin triples.

One can see that the dimensions of the subalgebras are equal and that bases $\{T_i\}, 
\{\tilde T^i\}$ in the subalgebras can be chosen so that
\be \langle T_i,T_j\rangle =0,\  \langle T_i,\tilde T^j\rangle =\langle \tilde T^j,T_i\rangle =\delta_i^j,\  
\langle \tilde T^i,\tilde T^j\rangle =0.\lbl{brackets}\ee
This canonical form of the bracket is invariant with respect to the transformations 
\be T_i'=T_k A^k_i,\ \tilde T^{'j}=(A^{-1})^j_k \tilde T^k. \lbl{tfnb}\ee
Due to the ad-invariance of $\langle .,.\rangle $ the algebraic structure of $\cd$ is
\[ [T_i,T_j]={f_{ij}}^k T_k,\ [\tilde T^i,\tilde T^j]={\tilde {f^{ij}}_k} \tilde T^k,\]
\be [T_i,\tilde T^j]={f_{ki}}^j \tilde T^k +{\tilde {f^{jk}}_i} T_k. \lbl{liebd}\ee

From the above given facts it is clear that the subalgebras $\cg$,$\tcg$ of the four--dimensional Drinfeld double 
are two--dimensional and surprisingly the Jacobi identities do not impose any condition on coefficients 
${ f_{ij}}^k,\tilde {f^{ij}}_k$ in this case.
Each of the subalgebras is solvable and due to the invariance of \rf{brackets}) w.r.t. \rf{tfnb}), the 
basis $\{ T_1,T_2 \}$ can be chosen so that the nontrivial Lie bracket in the first subalgebra is
\be [T_1,T_2]=nT_2 \lbl{lie1}\ee
where $n=0$ or 1. However, the Lie bracket in the second subalgebra in general cannot be written in a 
similar way without breaking the canonical form \rf{brackets}) of the bracket $\langle ,\rangle $ or 
the canonical form \rf{lie1}) of the subalgebra $\cg$. Nevertheless, we can use the transformations \rf{tfnb}) with
\be A  =  \left ( \begin{array}{cc} 
  1&a  \\
 0& b
  \end{array} \right ),\lbl{aspec}\ee
that preserve \rf{lie1}) to bring the Lie bracket of the second subalgebra to one of the following form
\be [\tilde T^1,\tilde T^2]=\beta \tilde T^2,\ \beta \in\real\ {\rm or} \ [\tilde T^1,\tilde T^2]=\tilde T^1.\lbl{lie2}\ee

In summary, {\it there are just four types of nonisomorphic four-dimensional Manin triples}. 
\\Abelian Manin triple:
\be [T_i,T_j]=0,\ [\tilde T^i,\tilde T^j]=0,\  [T_i,\tilde T^j]=0,\ i,j=1,2. \lbl{abel}\ee
\\Semiabelian Manin triple (only nontrivial brackets are displayed):
\be \label{sa} [\tilde T^1,\tilde T^2]=\tilde T^2,\  [T_2,\tilde T^1]=T_2,\ [T_2,\tilde T^2]=-T_1.\ee
\\Type A nonabelian Manin triple ($\beta\neq 0$):
\[ [T_1,T_2]=T_2,\  [\tilde T^1,\tilde T^2]=\beta\tilde T^2, \]
\be\label{tA} [T_1,\tilde T^2]=-\tilde T^2,\  [T_2,\tilde T^1]=\beta T_2,\  [T_2,\tilde T^2]=-\beta T_1+\tilde T^1. \ee
\\Type B nonabelian Manin triple:
\[ [T_1,T_2]=T_2,\  [\tilde T^1,\tilde T^2]=\tilde T^1, \]
\be \label{tB} [T_1,\tilde T^1]= T_2,\  [T_1,\tilde T^2]=-T_1-\tilde T^2,\  [T_2,\tilde T^2]=\tilde T^1. \ee
An interesting fact is that Drinfeld doubles corresponding to semiabelian Manin triple (\ref{sa}) and type B nonabelian Manin triple
(\ref{tB})
are the same, i.e. {\it these Manin triples are different decomposition into maximally isotropic subalgebras
of the same Lie algebra with the same invariant form}. The transformation   
of the dual basis between these decompositions is
\begin{eqnarray}
\nonumber X_1   = &  -\tilde{T}^1+\tilde{T}^2,  & 
 X_2  =  T_1+T_2, \\
\tilde{X}^1  =  & T_2,  & 
\tilde{X}^2  =  \tilde{T}^1,
\end{eqnarray}
where $(X_i,\tilde{X}^j)$ denotes the dual basis in the type B nonabelian Manin triple and
$(T_i,\tilde{T}^j)$ is the basis in the semiabelian Manin triple.
The other Manin triples specify the algebra of the Drinfeld double uniquely, i.e. there is
one connected and simply connected Drinfeld double to each of these Manin triples. 

\section{Dual sigma models}
Having all four--dimensional Drinfeld doubles we can construct the two--dimensional Poisson--Lie T--dual 
sigma models on them. The construction of the models is described in \cite{klse:dna} and \cite{kli:pltd}. 
The models have target spaces in the Lie groups $G$ and $\tilde G$ and are defined by the Lagrangians
\be {\cal L}= E_{ij}(g)(g^{-1}\partial_- g)^i(g^{-1}\partial_+ g)^j \ee
\be \tilde{\cal L}=\tilde E_{ij}(\tilde g)(\tilde g^{-1}\partial_- \tilde g)^i(\tilde g^{-1}\partial_+ \tilde g)^j \ee
where
\be E(g)=(a(g) + E(e)b(g))^{-1}E(e)d(g), \ee
$E(e)$ is a constant matrix
and $a(g),b(g),d(g)$ are $2\times 2$ submatrices of the adjoint representation of the group $G$ on $\cd$ in the basis $(T_i,\tilde T^j)$
\be Ad(g)^T  =  \left ( \begin{array}{cc} 
  a(g)&0  \\ b(g)&d(g)  \end{array} \right ). \ee
The matrix $\tilde E(\tilde g)$ is constructed analogously  with 
\be Ad(\tilde{g})^T  =  \left ( \begin{array}{cc} 
  \tilde{d}(\tilde{g})& \tilde{b}(\tilde{g})  \\ 0 & \tilde{a}(\tilde{g})  \end{array} \right ), \; \tilde E(\tilde e) = E(e)^{-1} =  
\left ( \begin{array}{cc} 
  x&y  \\ u&v  \end{array} \right ). \ee

Both equations of motion of the above given lagrangian systems can be reduced from equation of 
motion on the whole Drinfeld double, not depending on the choice of Manin triple:
\be\label{emkl}
\langle (\partial_{\pm} l) l^{-1}, {\cal E}^{\pm} \rangle =0,
\ee
where subspaces ${\cal E}^{+}= span(T^i+E^{ij}(e) \tilde{T}_j)$, ${\cal E}^{-}= span(T^i-E^{ji}(e) \tilde{T}_j)$ 
are orthogonal w.r.t. $\langle, \rangle$ and span the whole Lie algebra $D$.
One writes $l = g .\tilde{h}, 
\, g \in G,\, \tilde{h} \in \tilde{G}$ (such decomposition of group elements exists at least at the vicinity of the  unit  element) 
and eliminates $\tilde{h}$ from (\ref{emkl}), respectively $l = \tilde{g} .h, 
\, h \in G,\, \tilde{g} \in \tilde{G}$ and eliminates $h$ from (\ref{emkl}). The resulting equations of motion
for $g$, resp. $\tilde{g}$ are the equations of motion of the corresponding lagrangian system (see \cite{klse:dna}).

The corresponding models for the Drinfeld doubles \rf{abel})--\rf{tB}) are the following.
\\{\bf Abelian double:} The adjoint representations of the groups $G,\tilde G$ are trivial
so that
\be \tilde E(\tilde g)= \tilde E(e) = E(g)^{-1}=E(e)^{-1}, \ee
and the Lagrangians of the dual models are
\be {\cal L}=(xv-uy)^{-1}\left( v\,\partial_-\chi\partial_+\chi -y\,\partial_-\chi\partial_+\theta -u\,\partial_-
\theta\partial_+\chi +x\,\partial_-\theta\partial_+\theta \right),
\ee 
\be \tilde{\cal L}=x\,\partial_-\sigma\partial_+\sigma +y\,\partial_-\sigma\partial_+\rho +u\,\partial_-
\rho\partial_+\sigma +v\,\partial_-\rho\partial_+\rho.
\ee
\\{\bf Semiabelian double:} The adjoint representations of the groups $G,\tilde G$ are 
\[ Ad(g)^T=\left ( \matrix{ 1 & 0 & 0 & 0 \cr 0 & 1 & 0 & 0 \cr 0 & \theta  & 1 & 0 \cr -\theta  & 0 & 0 & 1 \cr  }\right),
\ \ \ Ad(\tilde{g})^T=\left ( \matrix{ 1 & 0 & 0 & 0 \cr -\rho  & e^{\sigma } & 0 & 0 \cr 0 & 0 & 1 & {\rho }
   {e^{-\sigma }} \cr 0 & 0 & 0 & e^{-\sigma } \cr  }\right)
\]
where $(\chi,\theta)$ and $(\sigma,\rho)$ are group coordinates of $G$ and $\tilde G$. The Lagrangians of the dual models are
\begin{eqnarray}\lbl{sa1}
\nonumber {\cal L} & = & \left(v\,x - u\,y - u\,\theta  + y\,\theta  +  \theta^2 \right)^{-1}\left[ 
v\,\partial_-\chi\partial_+\chi 
-(\theta+y)\,\partial_-\chi\partial_+\theta \right.
\\ & & \left.
+(\theta-u)\,\partial_-\theta\partial_+\chi
 +x\,\partial_-\theta\partial_+\theta \right], \\
\nonumber \tilde{\cal L} & = & \left(x-u\,\rho-y\,\rho+v\rho^2\right)\,\partial_-\sigma\partial_+\sigma 
+(y-v\rho)\,\partial_-\sigma\partial_+\rho \\
& & +(u-v\rho)\,\partial_-\rho\partial_+\sigma +v\,\partial_-\rho\partial_+\rho .
\lbl{sa2}\end{eqnarray}
Similarly one may use the other possible decomposition of the double into maximally isotropic subalgebras, i.e.  
{\bf type B nonabelian Manin triple}.
In this case the adjoint representations of the groups $G,\tilde G$ are 
\[ Ad(g)^T=\left( \matrix{ 1 & {\theta }{e^{-\chi }} & 0 & 0 \cr 0 & e^{-\chi } & 0 & 0 \cr 0 & -1 + 
   e^{-\chi } & 1 & 0 \cr -1 + e^{\chi } & \theta  - {\theta }{e^{-\chi }} & -\theta  & e^{\chi } \cr  }\right), \]
\[ Ad(\tilde{g})^T= \left( \matrix{ e^{-\rho } & -\sigma  & \sigma  - e^{\rho }\,\sigma  & -1 + e^{-\rho } \cr 0 & 1 & -1 + 
   e^{\rho } & 0 \cr 0 & 0 & e^{\rho } & 0 \cr 0 & 0 & e^{\rho }\,\sigma  & 1 \cr  }\right)
\]
and the Lagrangians of the dual models are
\begin{eqnarray} \lbl{tb1}
  {\cal L} & = & \left[v\,x +(e^\chi-1-y)(e^\chi-1+u) \right]^{-1}\left[ 
(v + u\,\theta  + y\,\theta  + 
      x\,{\theta }^2)\partial_-\chi\partial_+\chi 
 \right. \\ \nonumber 
 &  & +\left. \left( -1 + e^{\chi } - y - 
        x\,\theta  \right)\partial_-\chi\partial_+\theta
-\left( -1 + e^{\chi } + u + 
          x\,\theta  \right)\partial_-\theta\partial_+\chi
 +x\partial_-\theta\partial_+\theta \right], \\
\nonumber \tilde{\cal L} & = & \left[v\,x - u\,y + e^{\rho }\,\left( u - 2\,v\,x - y + 2\,u\,y \right)
      + e^{2\,\rho }\,\left( 1 + v\,x + y - 
     u\,\left( 1 + y \right)  \right)\right]^{-1}
\\
\nonumber & & \left[x\,\partial_-\sigma\partial_+\sigma 
+\left( v\,x - 
      e^{-\rho }\,v\,x +y + e^{-\rho }\,u\,y - u\,y - 
      x\,\sigma )\right) \,\partial_-\sigma\partial_+\rho 
\right. \\
\nonumber & &  \left.
 -\left( v\,x - 
      e^{-\rho }\,v\,x -u + e^{-\rho }\,u\,y - u\,y + 
      x\,\sigma \right)\,\partial_-\rho\partial_+\sigma \right. \\  
& & - \left. \left( u\,\sigma  + y\,\sigma   -v - 
        x\,{\sigma }^2 \right)\,\partial_-\rho\partial_+\rho \right].
\lbl{tb2}\end{eqnarray}
This model has the same equations of motion in the double (\ref{emkl}) as the previous one (up to
transformation of matrix $E(e)$ induced by the change of basis of algebra) and in this sense is equivalent 
to it.
\\{\bf Type A nonabelian doubles:}
The adjoint representations of the groups $G,\tilde G$ are 
\[ Ad(g)^T=\left ( \matrix{ 1 & \theta e^{-\chi } & 0 & 0 \cr 0 & e^{-\chi } & 0 & 0 \cr 0 & -\beta \,
       \theta e^{-\chi } & 1 & 0 \cr \beta \,\theta & \beta \,{\theta }^2
e^{-\chi } & -\theta  & e^{\chi } \cr  } \right), \]
\[ Ad(\tilde{g})^T=\left ( \matrix{ 1 & 0 & 0 & -\beta^{-1}\rho e^{-\sigma } \cr -\rho  & e^{\sigma } & 
{\beta }^{-1}{\rho } & \beta^{-1}\rho^2 e^{-\sigma } \cr 0 & 0 & 1 & {\rho }
   {e^{-\sigma }} \cr 0 & 0 & 0 & e^{-\sigma } \cr  }\right)
\]
where $\beta$ parametrizes different Drinfeld doubles. The Lagrangians of the dual models are
\begin{eqnarray}
\nonumber {\cal L} & = & \left(v\,x - u\,y + u\,\beta \,\theta  - y\,\beta \,\theta  + \beta^2\,\theta^2 \right)^{-1}
\left[ 
(v + u\,\theta  + y\,\theta  + x\,{\theta }^2)\partial_-\chi\partial_+\chi \right. \\
 & - & \left. \left( y + x\,\theta  - \beta \,\theta  \right)\partial_-\chi\partial_+\theta -\left( u + x\,\theta  + \beta \,\theta  \right)\partial_-\theta\partial_+\chi
 +x\partial_-\theta\partial_+\theta \right], \\
\nonumber \tilde{\cal L} & = & \left( {{\beta }^2 - 
      u\,\beta \,\rho  + y\,\beta \,\rho  + v\,x\,{\rho }^2 - u\,y\,{\rho }^2} \right)^{-1} \\
\nonumber & &
\left[{\beta }^2 \left( { x - u\,\rho  - y\,\rho  +
      v\,{\rho }^2}\right)\,\partial_-\sigma\partial_+\sigma \right. \\
\nonumber & & +\beta \,
      \left( y\,\beta  + v\,x\,\rho  - u\,y\,\rho  - v\,\beta \,\rho  \right)\,\partial_-\sigma\partial_+\rho  \\
& & \left .-\beta \,\left( - u\,\beta + v\,x\,\rho  - u\,y\,\rho  + 
          v\,\beta \,\rho  \right)\,\partial_-\rho\partial_+\sigma +v\,\beta^2\,\partial_-\rho\partial_+\rho \right] .
\end{eqnarray}
By rescaling $E(e)\mapsto E(e)/\beta,\ {\cal L}\mapsto {\cal L}\beta, \tilde {\cal L}\mapsto \tilde {\cal L}/\beta$ we obtain 
the $GL(2|\real)$ model found in \cite{kli:pltd}. It means that even though we have a one-parametric class of nonisomorphic 
Drinfeld doubles of type A the corresponding dual models are equivalent.

\section{Conclusions}
We have classified the four--dimensional Drinfeld doubles and the Poisson--Lie T--dual models on them.
The investigation of the Drinfeld doubles showed explicitly that neither the subalgebras $\cg$, $\tcg$ per se 
specify the Drinfeld double completely (viz. (\ref{tA}) vs. (\ref{tB})) nor the Drinfeld double fixes the
subalgebras $\cg$, $\tcg$ uniquely (viz. (\ref{sa}) and (\ref{tB})).
It turned out that besides the pair of dual models on $GL(2|\real)$ presented in \cite{kli:pltd} and the 
trivial abelian models, there exist two pairs of dual models  \rf{sa1}), \rf{sa2}) 
 and \rf{tb1}), \rf{tb2}) on the semiabelian double. This is the simplest (and the only one known to the authors) example of 
nontrivial modular space of $\sigma$-models mutually connected 
by Poisson--Lie T--duality transformation.

It would be very interesting to find whether any of the semiabelian 
or nonabelian models is integrable.

\end{document}